\documentclass{osa-article}
\usepackage[utf8]{inputenc}

\journal{josab}
\articletype{Research Article}

\begin{document}

\title{On anomalously large nano-scale heat transfer between metals}

\author{Carsten Henkel\authormark{*}
\&
Paul Philip Schmidt\authormark{}}

\address{\authormark{}Institute of Physics and Astronomy,
University of Potsdam,
\\
Karl-Liebknecht-Str. 24/25,
14476 Potsdam,
Germany}

\email{\authormark{*}henkel@uni-potsdam.de} 

\begin{abstract}
The non-contact heat transfer between two bodies is more
efficient than the Stefan--Boltzmann law, when the
distances are on the 
nanometer scale (shorter than Wien's wavelength), 
due to contributions of thermally excited near fields.
This is usually described in terms of the fluctuation electrodynamics
due to Rytov, Levin, and co-workers.
Recent experiments in the tip--plane geometry 
have reported ``giant'' heat currents between metallic (gold) objects,
exceeding 
even the expectations of Rytov theory. We discuss a simple model
that describes the distance dependence of the data and permits 
to compare to a plate--plate geometry, as in the proximity
(or Derjaguin) approximation.
We extract an area density of active channels which is of the
same order for the experiments performed by the groups of
Kittel (Oldenburg) and Reddy (Ann Arbor). It is argued that 
mechanisms that couple phonons to an oscillating
surface polarisation are likely to play a role. 
\end{abstract}

%
\section*{Introduction}

Thermal radiation provides a seminal topic in physics and
beyond~\cite{Chandrasekhar_Book}. 
Since Planck explained its spectrum, at the dawn of
quantum mechanics, it is clear that over
distances shorter than
the Wien wavelength (a few microns at room temperature), heat
transfer can be more efficient than the Stefan--Boltzmann law
because near fields are involved.
Fluctuating electromagnetic near fields 
are successfully described by a statistical approach to
macroscopic electrodynamics
(Maxwell--Langevin equations) developed by Rytov, Levin, and
co-workers~\cite{Rytov3_Book}. This theory has been 
developed in the field of nano-optics~\cite{Novotny_Book}
and is also relevant for colloidal 
forces~\cite{Parsegian_Book} and 
Casimir forces~\cite{BuhmannII_Book, VolokitinPersson_Book}.
Detailed studies of the radiative heat transfer at short
distances go back to Polder and van Hove~\cite{Polder_1971}
and others~\cite{Loomis_1994}. 
They have been motivated by experiments where a sharp tip is
scanned over a sample and the changes in tip temperature are
monitored to measure the heat current~\cite{Cravalho_1967, Williams_1986a, Dransfeld_1988}. 
This gives access to
surface properties that are complementary to electronic
tunnelling, with a somewhat worse spatial resolution. 

The framework of Rytov theory has been shaken, since data
are being reported that it cannot explain. Relevant examples
are the temperature dependence of the Casimir force between
metals\cite{BuhmannII_Book, VolokitinPersson_Book},
and the anomalous electric field noise observed in ion
traps~\cite{Brownnutt_2015}. 
In this paper, we address the heat transfer observed
at distances between $1$~and $10\,{\rm nm}$ by the groups
of A. Kittel (Oldenburg, Kloppstech \& al.~\cite{Kloppstech_2017})
and of P. Reddy (Ann Arbor, Cui \& al.~\cite{Cui_2017a}).
Comparison to
different models and numerical calculations based on Rytov
theory performed by the authors of 
Ref.~\cite{Kloppstech_2017} shows that the observed heat
transfer is orders of magnitude larger and that its distance
dependence is not well understood. In the experiment
of Cui \& al.~\cite{Cui_2017a}, the
results are interpreted in a different way: the authors apply 
a sequence of cleaning procedures to their tip and find a 
change in the dependence of heat conductivity on distance. The
``cleanest'' setting gives a heat current consistent with
Rytov theory, though close to the limit of detection sensitivity.
In a related experiment, the heat current through a 
molecular (single channel) contact was studied,
showing discrete steps given by the quantum of thermal 
conductance~\cite{Cui_2017b}. The spatial resolution of scanning
images obtained by the authors of Ref.\cite{Kloppstech_2017}
suggests, however, that the presence of adsorbed molecules
on the tip is unlikely in their setup. It is not surprising 
that the conflicting 
interpretations of these experiments have given rise to overt
discussions.

In this paper, we motivate a simple model to describe the
observed data and show that the 
experiments of Refs.\cite{Kloppstech_2017, Cui_2017a}
can be put into a common perspective (section~\ref{s:model}). 
The merit of the model is mainly its simplicity, it does not
yet provide a physical explanation. It generates heat current densities
that apply to a plate-plate geometry 
and are disturbingly close, however, despite obvious differences 
between the experimental setups. Fluctuation electrodynamics
predicts much smaller numbers (section~\ref{s:Rytov}).
This state of affairs invites questions
on the role of additional channels for heat transport
and how these could be ``activated'' at short distances. Obvious 
candidates for heat carriers are phonons, and indeed, the idea of
``phonon tunneling'' has become increasingly 
popular in recent years~\cite{Prunnila_2010, Altfeder_2010, Mahan_2011, Sellan_2012,Ezzahri_2014,Budaev_2015a}. At the
moment, these proposals do not reproduce the
observed distance dependence consistently, as pointed out
in Ref.\cite{Kloppstech_2017}. The basic mechanism for
phonon tunneling goes back to investigations in the 1980s
when anomalous infrared absorption has been
found in metallic nanoparticles, see 
Section~\ref{s:phonons}.

\section{Model for experimental data}
\label{s:model}

\begin{figure}[t]
\centerline{%
\includegraphics[width=6.0cm]{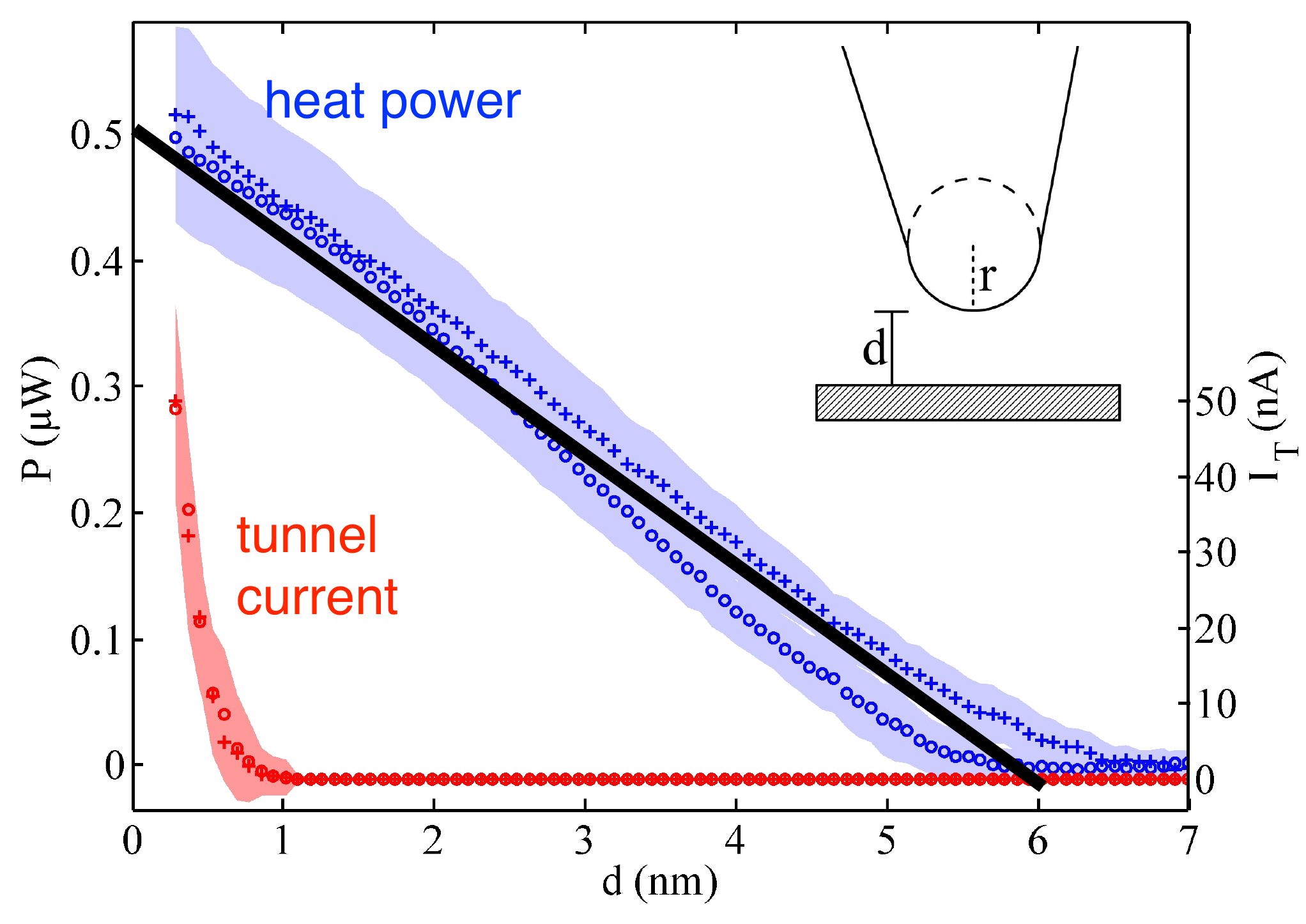}%
\includegraphics[width=6.0cm]{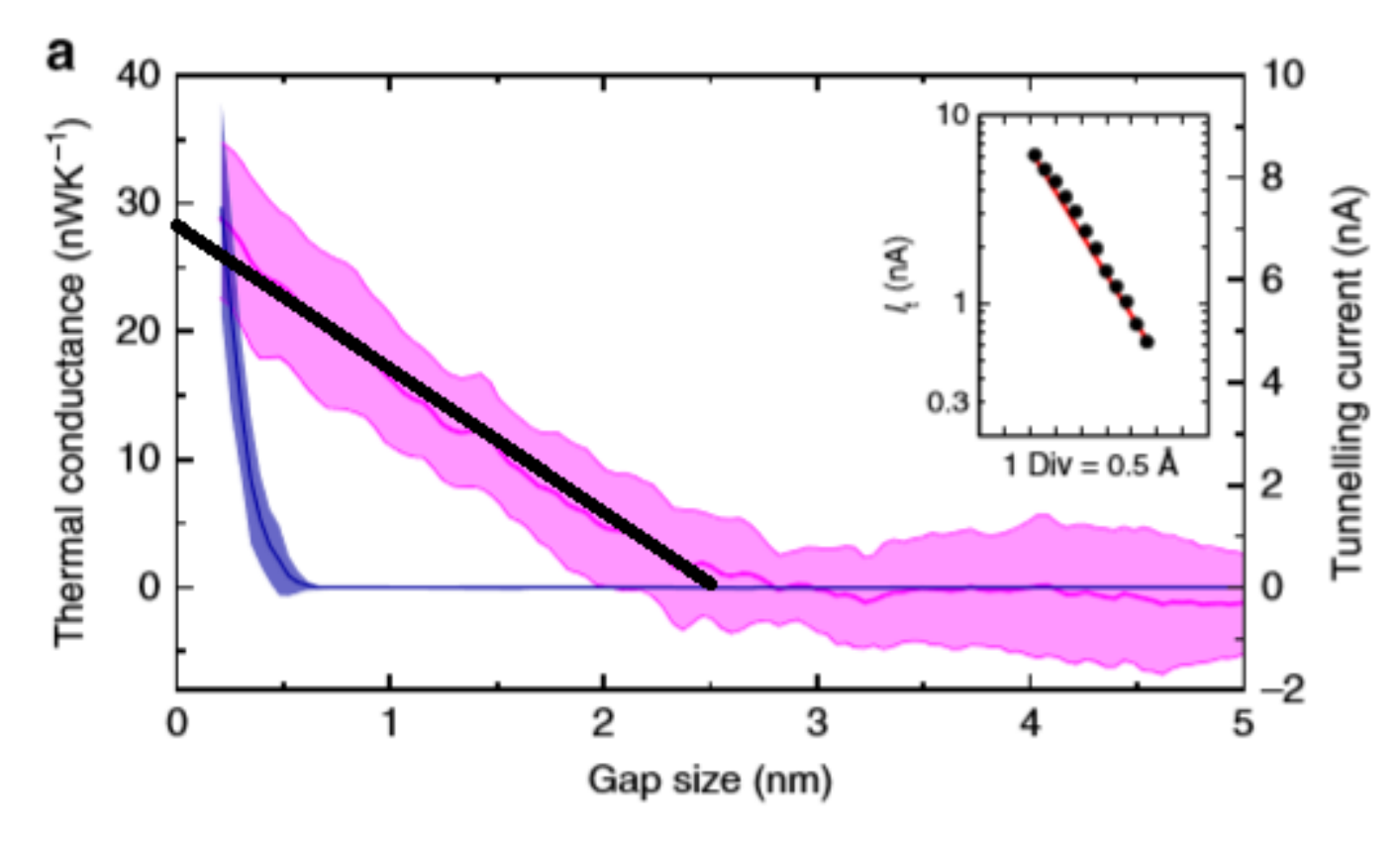}%
}
\caption[]{%
Experimental data on heat transfer between a tip
and a planar sample. The data were taken with thermal
scanning microscopes~\cite{Williams_1986a, Kloppstech_2015a}
that allow for a simultaneous measurement
of the electronic tunnelling current (onset at distances
shorter than $d \sim 0.5-1\,{\rm nm}$, right scale). 
The shaded areas illustrate the typical measurement 
uncertainties. Typical experimental parameters are collected
in Table~\ref{t:expt-data}. The thick solid lines give
the trend of our model~(\ref{eq:snap-in-model}).
\\
Left plot: heat power $P$, adapted from Fig.2 of 
Kloppstech et al., \emph{Nature Commun.} {\bf 8} (2017) 14475,
Ref.~\cite{Kloppstech_2017}, International License:
Creative Commons Attribution 4.0.
\\
Right plot: thermal conductance $G$ for a tip cleaned with an
organic solvent. Adapted from Fig.2(a) of 
Cui et al., \emph{Nature Commun.} {\bf 8} (2017) 14479,
Ref.~\cite{Cui_2017a}, International License:
Creative Commons Attribution 4.0.
}
\label{fig:two-data-sets}
\end{figure}

Two sets of experimental data on heat transport between 
a tip and a planar sample
are illustrated in Fig.\ref{fig:two-data-sets}. A striking
common 
feature is the relatively pronounced change in the distance
dependence: at large distance, the heat transfer is below
the detector noise, but upon approaching the sample, 
it increases linearly up to
the closest distances. 
An overview on the relevant experimental parameters is given
in Table~\ref{t:expt-data}. 
We note that the two setups have opposite `polarities':
Kloppstech \& al.\ \protect\cite{Kloppstech_2017} use a hotter tip, 
while Cui \& al.\ \protect\cite{Cui_2017a} use a colder one.
Table~\ref{t:expt-data} is based on data from 
Figs.\,2(a) and (b) in Ref.\protect\cite{Cui_2017a}, but not from
Fig.\,2(c) there where no significant linear increase is visible. 
In both experiments, the data do not suggest that electron tunneling
plays a significant role for the heat transfer, because the
heat and tunnel currents depend on distance in a very different
way.

\begin{table}[b]
\begin{center}
\begin{tabular}{l|ll}
\hline
& Kloppstech \& al. \cite{Kloppstech_2017} 
	& Cui \& al. \cite{Cui_2017a}
\\
\hline
tip temperature $T_t$ & $280\,{\rm K}$ & $303\,{\rm K}$
\\
sample temperature $T_s$ & $120\,{\rm K}$ & $343\,{\rm K}$
\\
tip radius $R$ & $30\,{\rm nm}$ & $150\,{\rm nm}$
\\
max. power $P$ & $0.5\,\mu{\rm W}$ & $1.2\,\mu{\rm W}$
\\
max. conductance $G$ 
		& $3\,{\rm nW/K}$ & $20 \ldots 30\,{\rm nW/K}$ 
\\
onset distance $d_c$ & $5 \ldots 6\,{\rm nm}$ & $2 \ldots 3\,{\rm nm}$
\\
\hline
\end{tabular}
\end{center}
\caption[]{Typical data for two experiments that measure
the heat transfer between a tip and a planar substrate.
To convert the
heat power $P$ to
the thermal conductance $G$, 
use the formula
$P = |T_t - T_s| G$
where $T_t$, $T_s$ are the tip and sample temperatures,
respectively. These temperatures are not measured right at
the tip apex, but further away along the tip shaft and below
the sample.
Note that the difference $T_t - T_s$ is not small.
}
\label{t:expt-data}
\end{table}

\begin{figure}[th]
\centerline{%
\includegraphics[width=5.0cm]{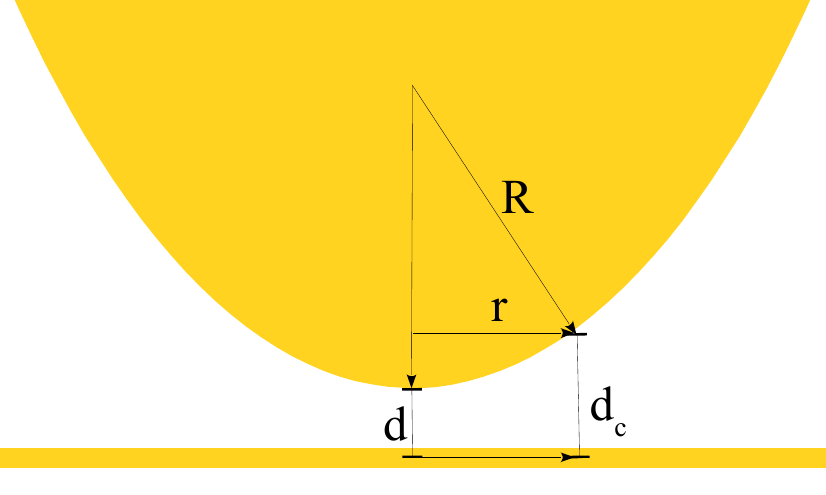}%
\raisebox{1.8cm}{\begin{minipage}{9cm}
\small
\begin{align}
A & \approx \pi r^2 &&\mbox{\qquad relevant tip area} \nonumber
\\
& = \pi ( R^2 - (R - d_c + d)^2) \nonumber
\\
& \approx 2\pi R (d_c - d) 
&&\mbox{\qquad if $R \gg d_c - d$}
\nonumber
\end{align}
\end{minipage}
}}
\caption[]{%
Sketch of the tip area that contributes to the anomalous heat
transfer.
}
\label{fig:sketch-tip-PFA}
\end{figure}

The linear relation between heat transfer and distance suggests
the following toy model that is related to the proximity force
(or Derjaguin) approximation~\cite{Parsegian_Book}.
Let us assume that as the distance $d$ between
two bodies drops below a critical value $d_c$ (much shorter
than the radius of curvature), a strong mechanism
of heat transport sets in and leads to a heat current $\dot q_c$
(a power per area). 
Assume further that heat is transferred 
in this way at a rate that does not depend on 
distance. Using elementary geometry and the ``blunt tip''
approximation (radius of curvature $R \gg d_c$, 
\emph{cf.} Fig.\ref{fig:sketch-tip-PFA}), 
we find that the ``active'' area 
grows linearly with $d_c - d$,
as one approaches the sample closer than the critical distance
$d_c$. 
The model can thus be summarized by the following formula for the
heat power 
[thick black lines in Fig.\ref{fig:two-data-sets}]
\begin{equation}
P = 2\pi R (d_c - d)\, \dot q_c
\,,
\qquad \mbox{for } d \le d_c
\,.
\label{eq:snap-in-model}
\end{equation}
Let us note that due to the large heat current, it is likely
that one gets a nontrivial temperature distribution
in the tip-sample contact area. Indeed, the hot tip generates
a small ``hot spot''~\cite{Altfeder_2010},
and the large current  $\dot q_c$ is ``draining heat'',
competing with thermal
conduction in the bulk of the material. If 
the tip diameter is comparable to the thermal phonon wavelength,
it is conducting significantly less heat than a bulk 
sample~\cite{Halbertal_2016}. This situation probably does not
apply to the experiments discussed here, because the phonon
wavelengths within the thermal spectrum are much shorter 
($< 1\,{\rm nm}$).

\begin{table}[b]
\begin{center}
\begin{tabular}{l|ll}
\hline
& Kloppstech \& al. \cite{Kloppstech_2017} 
	& Cui \& al. \cite{Cui_2017a}
\\
\hline
onset distance $d_c$ 
	& $5 \ldots 6\,{\rm nm}$ 
	& $2 \ldots 3\,{\rm nm}$
\\
saturated current $\dot q_c$ 
	& $4 \ldots 5 \cdot 10^{8}\,{\rm W/m}^2$ 
	& $5 \ldots 6 \cdot 10^{8}\,{\rm W/m}^2$
\\
channel density $n$ 
	& $1.6\,\text{\AA}^{-2}$ 
	& $4.5\,\text{\AA}^{-2}$
\\
Rytov prediction 
	& $2.3 \cdot 10^{6}\,{\rm W/m}^2$ 
	& $3.8 \cdot 10^{6}\,{\rm W/m}^2$
\\
\hline
\end{tabular}
\end{center}
\caption[]{Fit parameters for two nanoscale heat transport
experiments, based on Eq.(\ref{eq:snap-in-model}).
The channel density $n$ is computed by dividing the 
differential heat current density, 
$\dot q_c / |T_t - T_s|$,
by the effective thermal conductance quantum
$G_q = (\pi/6) (k_B^2/\hbar) \frac12 (T_t + T_s)$,
evaluated at the average temperature
[see Eq.(\ref{eq:upper-limit-Landauer})]. 
The Rytov prediction is the estimation
of fluctuation electrodynamics in the parallel
plate geometry~\cite{Polder_1971}. We have taken the 
leading order contribution  in the few-nm range, 
due to thermally excited magnetic near fields
[Eq.(\ref{eq:PvH-conductance})],
and have adapted the result to a large temperature difference.
}
\label{t:fit-data}
\end{table}

The simple model~(\ref{eq:snap-in-model})
describes the experimental data surprisingly well. 
The linear distance dependence of the heat power corresponds
indeed, by the proximity force approximation, 
to a distance-independent heat current density $\dot q_c$ 
in a plate-plate geometry. 
The parameters that can be extracted by fitting the model
to the data of Refs.\cite{Kloppstech_2017, Cui_2017a},
are given in Table~\ref{t:fit-data}. 
A striking observation 
is how close together are the values for the heat current
density $\dot q_c$,
despite the relatively large differences
in the measured heat power and the tip radius. 
We do not expect the difference in hot and cold
bodies to play a significant role here, as for the metallic 
materials used, the conditions for a thermal diode 
(strongly temperature-dependent thermal parameters) are not met.

We have also
computed the number $n$ of transport channels per unit area,
taking into account the quantum of thermal conductance $G_q$.
Due to the large differences in temperature, we have 
generalized the definition of $G_q$ by using the Landauer 
formula~\cite{BenAbdallah10,Biehs_2010}
for the heat current from the 
tip ($a$, $T_t$)
to the sample ($b$, $T_s$)
\begin{equation}
\dot q_{a \to b} =
n \int\limits_{0}^{\infty}\!\frac{ {\rm d}\omega }{ 2\pi }
\,
\frac{ k_B T_t\, |t_{ab}|^2 }{ \exp({ \hbar \omega / k_B T_t }) - 1 }
\label{eq:}
\end{equation}
The net current from tip to sample is given by the difference
$\dot q = 
|\dot q_{a \to b} - \dot q_{b \to a}|$.
If we assume that the
transmission is symmetric,
$|t_{ab}|^2 = |t_{ba}|^2$, as expected from reciprocity,
and limited by unity~\cite{BenAbdallah10,Biehs_2010}, 
then the maximum
heat current is
\begin{equation}
\dot q \le n 
\frac{ \pi }{ 12 } 
\frac{ k_B^2 }{ \hbar }
| T_t^2 - T_s^2 |
= 
n 
\underbrace{\frac{ \pi }{ 6 } 
\frac{ k_B^2 }{ \hbar }
\frac{ T_t + T_s }{2} }_{ G_q }
| T_t - T_s |
\label{eq:upper-limit-Landauer}
\end{equation}

Coming back to the experimental data extracted from this model
(Table~\ref{t:fit-data}),
we note that the channel densities $n$ differ only by 
a factor of $\sim 3$. In order of magnitude, they are larger
than one channel per unit cell (its cross section 
is $\sim 16.6\,\text{\AA}^2$ in gold), and this is only a lower
limit, provided the Landauer formalism 
applies~\cite{BenAbdallah10,Biehs_2010}. The channel density
would be higher if the transmission for many channels were
below unity, for example.
This illustrates 
that the heat transfer observed in these 
experiments is indeed a ``giant'' one~\cite{Kloppstech_2017}.
A question that is open so far is which mechanisms can possibly
lead to such an efficient coupling between two metallic bodies.

\section{Mechanisms for giant heat transfer}
\label{s:mechanisms}

\subsection{Fluctuation electrodynamics: magnetic near fields}
\label{s:Rytov}

The physical origin of the anomalously large heat transfer
has been addressed in previous work.
It has been pointed out that the standard fluctuation
electrodynamics of Rytov, Levin, and co-workers
falls short by a large factor~\cite{Kloppstech_2017}. 
Ignoring for the moment
the shape of the tip, we recall for simplicity an approximate
formula obtained by Polder and van Hove~\cite{Polder_1971}.
It turns out that at distances
in the $10\,{\rm nm}$ range,
the main contribution to the heat transfer between metals
comes from the ``magnetic near field'', i.e., evanescent
waves in the TE-polarization. It is \emph{not} dominated by 
surface plasmons, since these appear at frequencies way above
the room temperature Planck spectrum and
they have only a small mode density
in the relevant infrared range.
Ref.~\cite{Polder_1971} focuses
on the differential heat conductance per unit area
and finds the following approximation for the heat current
\begin{equation}
\dot q_{{\rm TE,\,evan}} 
\simeq
\frac{ 0.574 }{ 4\pi } \, \frac{ \mu_0 k_B^3 }{ \hbar^2 }
\sigma T^2
|T_t - T_s|
\qquad \mbox{for}\quad 
|T_t - T_s| \ll T_t, T_s
\label{eq:PvH-conductance}
\end{equation}
where 
$\mu_0$ is the magnetic constant,
$\sigma$ the DC conductivity (in SI units), 
and 
$T = \frac12 (T_t + T_s)$ the average temperature. This
formula applies when the distance is smaller than the
typical diffusion length
for thermal
magnetic fields, $\ell_{T} \sim 
[\hbar / ( k_B T \mu_0 \sigma )]^{1/2}$
(also known as skin depth). For the range
of temperatures considered here, $\ell_{T} \sim 15 \ldots 9\,{\rm nm}$.

If this estimate is extrapolated to a large temperature difference
(ignoring the change in conductivity), one gets instead of 
Eq.(\ref{eq:upper-limit-Landauer}) a heat current proportional
to $|T_t^3 - T_s^3|$. The corresponding predictions are included
in Table~2 and are two orders of magnitude lower than the values
extracted from the experiments. This discrepancy is even larger
($\sim 10^3$)
in fluctuation electrodynamics calculations 
based on the actual tip geometry~\cite{Kloppstech_2017}.

\subsection{``Activated'' phonons}
\label{s:phonons}

Attractive alternatives to fluctuation electrodynamics 
are mechanisms related to phonons.
Indeed, in metals at not too low temperatures, phonons already
give the largest contribution to the specific heat.
They are 
less effective than electrons for heat conduction,
however,
because the speed of sound is much smaller than the Fermi velocity.
In addition, phonons are not ``active'' in terms of coupling to
electric or magnetic fields: 
due to the fcc symmetry of the unit cell, we do not expect any
optical phonons in bulk gold, and the acoustic branch is not
accompanied by an oscillating polarisation. 
(This becomes different for 
piezoelectric materials, as pointed out in Ref.\cite{Prunnila_2010}.)
Several papers have suggested mechanisms to couple phonons
between two vacuum-separated bodies, also 
called ``phonon tunneling''. This is based, for example,
on the van der Waals coupling between surface 
oscillations~\cite{Sellan_2012,Ezzahri_2014,Budaev_2015a}
which has also been considered for thermal fluctuations
of the van der Waals force on a particle near a 
surface~\cite{Henkel_1999b}.
The drawback is that these proposals lead to a distance dependence 
that differs significantly from what is observed in 
Fig.\ref{fig:two-data-sets}~\cite{Kloppstech_2017}.

One should note here that the tips in the experiments
are very likely made of amorphous rather than crystalline gold.
And even in the case of flat surfaces (surface scans
in the experiments of Refs.\cite{Kloppstech_2017, Altfeder_2010} 
show atom-scale
steps, for example), it is well known that a gold surface
is undergoing a reconstruction over a depth of a few layers. 
This possibly
leads to optical surface phonons, featuring a net dipole moment. 
Hence
we may expect an oscillating surface polarization that 
is excited when bulk phonons scatter from the surface.
It is noteworthy that this has been discussed in the 1980s
when an anomalously large infrared absorption has been
observed in metallic nanoparticles (see references 
in Ref.\cite{Monreal_1985}). An interesting observation
has been made by Andersson and co-workers~\cite{Andersson_1984}:
they considered a ``jellium'' half-space and
showed that the electron density response
to an applied electric field~\cite{Feibelman_1982,Lang_1973} 
oscillates with the depth into
the metal, similar to Friedel oscillations. 
This gives a net charge imbalance in the first few layers.
The associated forces on the atomic cores drive an
optical surface phonon mode that dissipates by radiating
bulk phonons. In reverse, this mechanism may ``activate''
bulk phonons by generating a surface polarization that couples
electro\-statically to another body nearby. A similar idea
based on the dynamic behaviour of the image charge induced
in a metal surface has been put forward in Ref.\cite{Altfeder_2010}.
The distance dependence of the corresponding heat transfer
will be dealt with in a separate theory paper.

\section*{Conclusion}

We have proposed a simple model that permits to appreciate
the experimental data on non-radiative heat transfer observed
with scanning thermal microscopes on the sub-$10\,{\rm nm}$
scale~\cite{Cui_2017a, Kloppstech_2017}. Although the two setups
differ in their parameters and the two groups give different
interpretations of their respective observations, the model
puts the ``raw data'' into a common perspective, with surprisingly
close values for the heat conductance per area. The giant or
anomalous character of the observations is underlined by the 
large area density of heat-carrying channels that the model
predicts within the Landauer formalism. 

It is not likely that the tunnelling of electrons contributes
significantly to the observed heat transfer. Just by looking at the
distance dependence of the data for the tunnelling current and the 
heat flux (Fig.\protect\ref{fig:two-data-sets}), 
it seems rather clear that no significant electron transfer
occurs in the 2--5\,nm range in both experiments. Kloppstech \& 
al. \protect\cite{Kloppstech_2017} also do not mention electronic heat 
transfer among their list of candidate mechanisms.

There is a difference in the critical distance $d_c$ between the 
two setups which suggests that this parameter is not a fundamental
quantity. It may depend on local phenomena at the tip like
elastic deformations or migrating atoms, and this is likely to
depend on tip preparation (pulling, annealing etc.). The resolution
of the scanning images
shown in the supplementary material by Kloppstech \& al 
\protect\cite{Kloppstech_2017} does not suggest a contamination 
by molecules as long as a few nanometers.

We hope that our
considerations inspire further experiments and foster 
the search for a theoretical explanation
-- which must go beyond traditional Rytov--Levin fluctuation
electrodynamics. We have argued that genuine surface properties
like optical surface phonons that have been put forward in
the 1980s already~\cite{Andersson_1984, Monreal_1985}, 
may ``catalyze'' heat transport.
It can thus be expected that in any meaningful model for this
thermal anomaly, 
two significant differences with respect to
standard fluctuation electrodynamics appear: (i) due
to the large heat current, one probably needs to construct
a self-consistent description of the temperature profile
(see, e.g., Ref.~\cite{Messina_2016}),
and (ii) the models cannot be built from bulk material properties
and the response to electromagnetic fields is determined by
genuine surface properties~\cite{BedeauxVlieger_Book,Feibelman_1982}.

\section*{Funding}
Deutsche Forschungsgemeinschaft (DFG) (Schm 1049/7-1, Fo 703/2-1)

\section*{Acknowledgments}

We appreciated constructive remarks from anonymous referees.
We acknowledge Math.-Nat. Fakult\"at (Universit\"at Potsdam) 
for providing travel support to PPS.


%
%

\end{document}